\documentstyle[12pt]{article} \def\dspace{\baselineskip = .30in}

\begin{document}

\title{ Smooth Hybrid Inflation}

\author {{\bf G. Lazarides and C. Panagiotakopoulos}\\Physics
Division\\School of Technology\\University of
Thessaloniki\\Thessaloniki, Greece}
\date{ }
\maketitle

\dspace
\centerline{\bf Abstract}
\vspace{.2in}
\par
We propose a variant of hybrid inflation which is applicable in a wide
class of supersymmetric grand unified models.
The observed temperature perturbations of cosmic
background radiation can be reproduced with natural values of the
parameters of the theory and a grand unification scale which is
consistent with the unification of the minimal supersymmetric
standard model gauge couplings as measured at
LEP. The termination of
inflation is smooth and does not produce any topological defects.
Finally, we present a specific supersymmetric model where our smooth
hybrid inflationary scenario is realized.
\newpage
\par
The usual realizations of the new and the chaotic inflationary
scenarios$^{(1)}$
invoke a very weakly coupled gauge singlet scalar field known as
inflaton. The extremely weak couplings of this  field, which are
necessitated by the smallness of  the observed temperature
fluctuations in the cosmic background radiation (CBR) are certainly
unsatisfactory because they create an extra `hierarchy' or
`naturalness' problem. Recently, Linde$^{(2)}$ has proposed a new realization
of the chaotic inflationary scenario based on a coupled system of two
scalar fields one of which may not be a gauge singlet. The great
advantage
of this hybrid inflationary model is that it produces the observed
temperature fluctuations in the CBR with natural values of the
coupling constants. However, inflation terminates abruptly and is
followed by a `waterfall' regime during which topological defects can
be easily produced. Consequently, in the cases where these defects
include superheavy magnetic monopoles or domain walls we end up with an
unacceptable
cosmology. Of course, this is not the case with the original hybrid
inflationary model$^{(2)}$.
Hybrid inflation has been adjusted$^{(3)}$ so
that it becomes applicable in a wide class of supersymmetric (SUSY)
grand unified theories (GUTs) based on semi-simple gauge groups. The
magnitude of the temperature fluctuations in CBR turns out to be
directly related to the grand unification scale $M_X$. However, values
of $M_X$ consistent with the unification of the gauge
couplings$^{(4)}$ of the
minimal supersymmetric standard model (MSSM), which is favored by
recent LEP  data, tend to give values of the temperature fluctuations
in CBR  considerably higher than the values allowed by observations.
Also, the scheme heavily relies on radiative corrections.

\par
In this paper we propose a variant of Linde's potential which can also
be derived in a wide class of SUSY GUTs based on semi-simple gauge
groups. Instead of the renormalizable trilinear coupling in the
superpotential we utilize the first non-renormalizable contribution.
All our analysis is done at the tree-level and no radiative
corrections are needed. Although we get only a  slight variation of
Linde's potential, the cosmological scenario obtained is drastically
different. Already since the beginning of inflation, the system follows
a particular valley of minima which leads to a particular point of the
vacuum manifold. So our inflationary scenario does not lead to
production of topological defects. Also the termination of inflation
is not as abrupt as in the hybrid case. It is quite smooth and
resembles more the cases of new or chaotic inflation. The main
advantage of our smooth inflationary scenario is that the measured
value of the temperature fluctuations of CBR can be reproduced with
natural values of the  parameters and with a GUT scale, $M_X$,
consistent with the
unification of the MSSM gauge couplings. It is
also remarkable that the scale controlling the non-renormalizable
terms in the superpotential turns out to be of order $10^{18}$ GeV.
The spectral index of density fluctuations is close to unity.

\par
We consider a SUSY GUT based
on a (semi-simple) gauge group $G$ of rank $\geq$5. $G$ breaks
spontaneously directly to the standard model (SM) group $G_S$
at a scale $M_X \sim 10^{16}$ GeV. We assume that below $M_X$
the only SM non-singlet states of the theory are the
usual MSSM states in order for the successful  MSSM predictions for
$sin^2 \theta_w$ and $\alpha_s$ to be retained. The theory could also
possess some global symmetries. The symmetry breaking of $G$ to $G_S$ is
obtained through a superpotential which includes the terms
\begin{equation}
W = s (- \mu^2 + \frac {(\phi \bar{\phi})^2} {M^2}) .
\end{equation}
Here $\phi,\bar{\phi}$ is a conjugate pair of left-handed SM singlet
superfields which belong to non-trivial representations of the gauge
group G
and reduce its rank by their vacuum expectation values (vevs), s is a
gauge singlet left-handed superfield, $\mu$ is a superheavy mass scale
related to $M_X$, whereas $M$ is a mass scale of the order of the
`compactification' scale $M_c \sim 10^{18}$ GeV which controls the
non-renormalizable terms in the superpotential of the theory. The
superpotential terms in eq.(1) are the dominant couplings involving
the superfields s, $\phi, \bar{\phi}$ consistent with a continuous
R-symmetry under which $W \rightarrow e^{i \theta} W, s \rightarrow
e^{i \theta}s, \phi \bar{\phi} \rightarrow \phi \bar{\phi}$ and a
discrete symmetry under which $\phi \bar{\phi}$ changes sign.
Moreover, we assume that the presence of other SM singlets in the
theory does not affect the superpotential in eq. (1) which is
responsible for the non-zero vevs  of the fields $\phi, \bar{\phi}$.
This, if not automatic, is very easily achieved in the case of
semi-simple gauge groups. The potential obtained from $W$ in eq.(1), in
the supersymmetric limit, is
\begin{equation}
V = | \mu^2 - \frac {(\phi \bar{\phi})^2}{M^2}|^2 + 4|s|^2 \frac
{|\phi|^2|\bar{\phi}|^2}{M^4} (|\phi|^2 + |\bar{\phi}|^2) + D-terms,
\end{equation}
where the scalar components of the superfields are denoted by the same
symbols as the corresponding superfields. Vanishing of the D-terms is
achieved with $|\bar{\phi}|=|\phi|$ (D-flatness condition). The
supersymmetric vacuum
\begin{equation}
<s>=0, \> \> <\phi><\bar{\phi}> = \pm \mu M, \> \>
|<\bar{\phi}>|=|<\phi>|
\end{equation}
lies on the D-flat direction $\bar{\phi}^*=\pm \phi$. Restricting
ourselves to this particular direction and performing appropriate
gauge, discrete and R-transformations we can bring the complex
s,$\> \phi,\bar{\phi}$ fields on the real axis, i.e,  $s \equiv
\frac{\sigma}{\sqrt{2}}, \> \> \bar{\phi}=\phi \equiv
\frac{1}{2}\chi$, where $\sigma$ and $\chi$ are real scalar fields.
The potential in eq.(2) then takes the form
\begin{equation}
V(\chi, \sigma) = (\mu^2 - \frac{\chi^4}{16M^2})^2 + \frac {\chi^6
\sigma^2} {16M^4}
\end{equation}
and the supersymmetric minima correspond to
\begin{equation}
|<\chi>| = 2 (\mu M)^{1/2}, \> \> <\sigma> = 0 .
\end{equation}
The mass acquired by the gauge bosons is $M_X = g (\mu M)^{1/2}$,where
$g$ is the GUT gauge coupling. The vacuum manifold of the theory, which
is obtained from the minimum in eq.(5) by performing all possible
gauge and global transformations, may have non-trivial homotopical
properties in which case the theory predicts the existence of
topological defects.

\par
Let us now take a closer look at the potential in eq. (4). For any
fixed value of $\sigma$, this potential, as a function of $\chi^2$,
has a local maximum at $\chi^2=0$ and an absolute minimum lying at
\begin{equation}
\chi^2 \simeq \frac{4}{3} \frac{\mu^2M^2}{\sigma^2} , \> \> for \> \>
\sigma^2 >> \mu M,
\end{equation}
and
$\chi^2 \simeq 4 \mu M$, for $ \sigma^2 << \mu M$.
The value of the potential along the maxima at $\chi^2 = 0$ is
constant, $V_{max} (\chi^2 = 0) = \mu^4$, whereas along the valley of
minima, for $\sigma^2 >> \mu M$, is $V_{min} (\sigma) \simeq \mu^4
[1-(2/27)(\mu^2M^2/\sigma^4)]$.

\par
We are now ready to turn to the discussion of the cosmological
evolution of this system. We assume that after `compactification' at a
cosmic time $t_c \sim (\frac{3}{8 \pi})^{1/2} \frac{M_P}{M^2_c}$,
where $M_P =1.2 \times 10^{19} $GeV is the Planck mass, the universe
emerges with energy density of order $M^4_c$. We then, as usual, follow
the development of  a region of size $\ell_c \sim t_c$ where the
scalar fields $\chi$ and $\sigma$ happen to be almost uniform with
$|\sigma| >>|\chi|$. Under these circumstances, one can easily see
that initially the last term in eq.(4) is the dominant contribution to
the potential energy of the system. Thus, the initial equations of
motion for the $\chi$ and $\sigma$ fields read
\begin{equation}
\ddot {\chi} + 3 H \dot {\chi} + \frac {3 \sigma^2 \chi^5}{8M^4} \simeq
0
\end{equation}
and
\begin{equation}
\ddot {\sigma} + 3H \dot {\sigma} + \frac {\chi^6 \sigma}{8M^4} \simeq
0,
\end{equation}
with $H$ being the Hubble parameter and dots denoting derivatives with
respect to cosmic time. Assuming for the moment that
$\sigma$ remains approximately constant, we see that the frequency of
oscillation of the $\chi$ field, which is of order
$\chi^2_m|\sigma|/M^2$ ($\chi_m$ is the amplitude of the oscillations
of $\chi$), is much greater that the Hubble parameter $H \simeq
(\pi/6)^{1/2} (\chi^3_m |\sigma|/M_P M^2$), for $\chi_m << M_P$. So
$\chi$ initially performs damped oscillations over the maximum at
$\chi=0$. For $|\sigma|>>M_P$, the `frequency' of oscillation of the
$\sigma$ field, which is of order $\chi^3_m/M^2$, is much smaller than
$H$.
Thus, our hypothesis that $\sigma$ does not change much  is justified.
When the amplitude of the $\chi$ field drops to about $(\mu^2
M^2/|\sigma|)^{1/3}$, the $\mu^4$ term dominates the potential in eq.
(4) and the Hubble parameter becomes approximately constant and equal
to $H=(8 \pi/3)^{1/2} \mu^2/M_P$ and remains so thereafter till the
end of  inflation. However, the form of the eqs.(7) and  (8) still holds
till $\chi_m$ drops to about $\mu M/|\sigma|$. The `frequency' of the
$\sigma$ field remains much smaller than $H$, for
$(\mu^2 M^2/|\sigma|)^{1/3} \stackrel {_>}{_\sim} \chi_m \stackrel
{_>}{_\sim} \mu M/ |\sigma| \> \> (|\sigma| >> M_P)$ and eq.(8) reduces to
\begin{equation}
3 H \dot{\sigma} + \frac {\chi^6 \sigma}{8M^4} \simeq 0.
\end{equation}
As is easily shown, $\sigma$ again remains almost constant within one
expansion time, i.e., $|\Delta \sigma/ \sigma| << \chi^6_m M^2_P/
\mu^4 M^4 \leq (M_P/ |\sigma|)^2 << 1$.
The frequency of the $\chi$ field $\sim \chi^2_m |\sigma| / M^2$
remains greater than $H$, for $\chi_m \stackrel {_>}{_\sim} \mu M/
(M_P|\sigma|)^{1/2} \geq \mu M /|\sigma| $ and $\chi$ still performs
damped oscillations. For smaller values of the $\chi \>$  field $(\mu M
/(M_P|\sigma|)^{1/2} \stackrel {_>}{_\sim} \chi \stackrel {_>}{_\sim}
\mu M /|\sigma|)$, the Hubble parameter overtakes and $\chi$ enters
into a `slow roll-over'  regime controlled by the equation
\begin{equation}
3 H \dot {\chi} + \frac {3 \sigma^2 \chi^5} {8M^4} \simeq 0.
\end{equation}
Eqs.(9) and (10) then imply that $3 \sigma^2 - \chi^2 \simeq$ constant.
Thus, $\chi$ keeps dropping without any essential change in $\sigma$
and, as one can deduce from eq.(10), reaches a value $\sim \mu M
/|\sigma|$ in a cosmic time interval $\Delta t \sim  (\sigma /M_P)^2
H^ {-1} >> H^{-1}$. After that, the $\chi^4$ term in eq.(4) comes into play
and the equation of motion of the $\chi$ field becomes
\begin{equation}
\delta \ddot{\chi} + 3 H \delta \dot{\chi} + m^2_\chi (\sigma) \delta
\chi \simeq 0,
\end{equation}
where $\delta \chi$ is the deviation of $\chi$ from the minima in
eq.(6)
and $m^2_\chi (\sigma) \simeq 4 \mu^4/3 \sigma^2$ is the $\sigma$
-dependent mass squared of the $\chi$  field along the valley of these
minima. For $|\sigma| >> M_P, \> \> m_\chi(\sigma) << H $ and eq. (11)
reduces to
\begin{equation}
3 H \delta \dot {\chi} + m^2_\chi (\sigma) \delta \chi \simeq 0.
\end{equation}
Thus, $\chi$ reaches the minimum in a time interval
\begin{equation}
\Delta t \sim 6 \pi (\frac{\sigma}{M_P})^2 H^{-1}
\end{equation}
within which the $\sigma$ field stays essentially unaltered as one can
see from eq.(9). To summarize, so far, we have seen that within a time
interval given in eq.(13) the $\chi$ field falls into the valley of
minima in eq.(6) and relaxes at the bottom of this valley whereas the
$\sigma$ field still remains unchanged and much greater than $M_P$. Of
course, this does not automatically mean that $\chi$ follows the
valley of minima at subsequent times but, as we shall soon see, it
turns out that it actually does.

\par
One could also reach the same conclusion starting with field values
considerably smaller than $M_P$. This requires that the initial
energy density be $\sim \mu^4$ for $\sigma$ to be slowly varying.
Consider the situation where the initial value of $\sigma$ is close to
$M$. It is then easy to show that, for initial values of $\chi$ much
smaller than $(4 \sqrt{3 \pi} \frac {(\mu M)^{1/2}}{M_P})^{1/3} (\mu
M)^{1/2}$ but still greater than the value given in eq.(6), the $\chi$
field performs damped oscillations and ends up  at the bottom of the
valley of minima with $\sigma$ remaining essentially unaltered.

\par
To discuss the further evolution of the system let us, for the moment,
suppose that $\chi$, at subsequent times, follows the valley of minima
in eq. (6). Eq. (9) then reads
\begin{equation}
3 H \dot{\sigma} + \frac {8 \mu^6 M^2} {27 \sigma^5} \simeq 0 ,
\end{equation}
which gives
\begin{equation}
\sigma \simeq (\frac {2 \mu M M_P}{3 \sqrt{2 \pi}})^{1/3}
N^{1/6}(\sigma),
\end{equation}
where $N(\sigma) \equiv H \Delta t$ is the number of e-foldings from
the moment at which the $\sigma$ field has the value $\sigma$ till the
end of inflation.
Inflation ends when the `frequency' of the $\sigma$ field becomes
smaller than $3H/2$, i.e., at
\begin{equation}
\sigma \simeq \sigma_o \equiv (\frac {2M_P}{9 \sqrt {\pi}(\mu
M)^{1/2}})^{1/3} (\mu M)^{1/2} \stackrel {_>}{_\sim} (\mu M)^{1/2} .
\end{equation}
For $|\sigma|>M_P$, the evolution of $\chi$ is governed by eq.(12)
and the time needed to reach its minimun (see eq.(13)) is much smaller
than $N(\sigma) H^{-1}$ from eq.(15) which is the time required for
$\sigma$ to change significantly. Also the quantum fluctuations of
$\chi$  in de Sitter space, $\delta \chi \simeq H/2 \pi$, are not strong
enough to surpass the barrier at $\chi = 0$ if $\sigma << \sqrt {2
\pi} (M/ \mu)M_P$. For $\sigma \stackrel {_<} {_\sim} M_P, \> \>
m_\chi
\stackrel {_>}{_\sim} H$, the quantum fluctuations of $\chi$ stop and
we go back to eq.(11) which describes damped oscillations about the
minimum in eq.(6) with  damping time $H^{-1}$ much smaller than
$N(\sigma) H^{-1}$. Thus, we conclude that our starting assumption
that $\chi$ follows the valley of minima is correct. Note that, due to
supersymmetry breaking, the $\chi$ and $\sigma$ fields acquire
additional masses of order $M_s \sim 1$ TeV. It is easy to check that
this mass contribution is never important for the evolution of the
$\chi$ field. However, it plays a role for the $\sigma$ field if its
initial value happens to be greater than $\sigma_s \equiv
(2/\sqrt{6}) (M/M_s)^{1/3} \mu $.
This introduces only unimportant complications to the above
discussion which we will skip. In any case, as soon as $\sigma$ drops
below $\sigma_s$ its SUSY breaking mass becomes subdominant.

\par
The contribution of the scalar metric perturbation to the
microwave background quadrupole anisotropy (scalar Sachs-Wolfe effect)
is given$^{(5)}$ by
\begin{equation}
(\frac {\Delta T}{T})_S \simeq (\frac {32 \pi}{45})^{1/2} \frac
{V^{3/2}}{M^3_P (\partial V/ \partial \sigma)} |_{k \sim H} = 9 ( \frac
{\pi} {10})^{1/2} \frac {\sigma^5}{M^3_P M^2}|_{k \sim H} ,
\end{equation}
where the right-hand side is evaluated at the value of the $\sigma$
field where  the lenght scale $k^{-1}$, which corresponds to the present
horizon size, crossed outside the de Sitter  horizon during  inflation. The
derivative $(\partial V/ \partial \sigma)$ is, of course, calculated
on the valley of minima of eq.(6) and substituting $\sigma$ in terms
of $N(\sigma)$ from eq.(15) and $\mu$ in terms of $M_X$, we obtain
\begin{equation}
(\frac {\Delta T} {T})_S \simeq 5^{-1/2} 6^{1/3} \pi^{-1/3}
N^{5/6}_H g^{-10/3} M_X^{10/3} M_P^{-4/3} M^{-2},
\end{equation}
where $N_H$ is the number of e-foldings of the present horizon size
during inflation. Taking $N_H=60$, $M_X = 2 \times 10^{16}$ GeV,  $g=0.7$
(consistent with the MSSM unification) and $(\Delta T/T) \simeq
5 \times 10^{-6}$ from COBE, we get $M \simeq 9.4 \times 10^{17} $GeV and $\mu
\simeq 8.7 \times 10^{14}$GeV.
We ignored the gravitational wave contribution $(\frac {\Delta
T}{T})_T$ to the quadrupole
anisotropy
since it turns out to be utterly negligible relative to the
scalar component:
\begin{equation}
(\frac{\Delta T}{T})_T \simeq 0.78 \frac {V^{1/2}}{M_P^2} \simeq 4.1
\times
10^{-9}.
\end{equation}
The amplitude of the density fluctuations on a given length
scale $k^{-1}$ as
this scale crosses inside the postinflationary horizon is proportional
to $N^{5/6}_{k^{-1}} \> \simeq \>$ \hfil\break
$N^{5/6}_H [k^{-1} (Mpc)/10^4]^{5/6 N_H}$,
where $N_{k^{-1}}$ is the number of e-foldings of this scale during
inflation (see eq.(18)). The spectral index is then given by
\begin{equation}
n = 1 - \frac{5}{3N_H} \simeq 0.97 \> \> (for \> \> N_H \simeq 60),
\end{equation}
which is very close to the Harrison-Zeldovich value $(n=1)$ and lies in
the central range of values preferred by observations. We can also
estimate the value of the $\sigma$ field at the end of inflation (see
eq.(16)) $\sigma_o \simeq 1.1 \times 10^{17}$ GeV $\simeq 5.4 M_X$ and its
value when the present horizon size crossed outside the inflationary
horizon (see eq.(15)) $\sigma_H \simeq (9 N_H /2)^{1/6} \sigma_o
\simeq 2.54 \sigma_o \simeq 2.7 \times 10^{17}$GeV. The total number of
e-foldings during inflation (see eq.(15)) turns out to be greater than
$10^{12}$, for initial $|\sigma| \geq M_P$, and SUSY breaking becomes
irrelevant if the initial value of the $\sigma$ field does not exceed
$\sigma_s \simeq 5.8 M_P$ (for $M_s \simeq 1$ TeV).

\par
The termination of inflation is not as abrupt as in the hybrid
inflationary scenario where the slow roll-over regime is followed by
the `waterfall' . In our case, after the end of inflation the $\sigma$
and $\chi$ fields enter smoothly into an oscillatory phase about the
global supersymmetric minimum of the potential in eq.(5) as in the
`new' inflationary scenario. The frequency of the oscillating fields
is $m_\sigma = m_\chi= 2 \sqrt{2} (\mu/M)^{1/2} \mu \simeq 7.6 \times
10^{13}$ GeV.
These fields should eventually decay into lighter particles and
`reheat' the universe so that its baryon asymmetry can be subsequently
produced. We will postpone until later the discussion of these important
issues because it depends strongly on the details of the particle
physics model one adopts.

\par
Our inflationary scenario, in contrast to the hybrid one, produces no
topological defects in the universe. This is because the $\sigma$
and $\chi$ fields during inflation follow everywhere the same valley
of minima which leads to a specific point of the vacuum manifold of
the theory. As we discussed earlier, the $\chi$ field stays close to
the bottom of the valley and is unable to fluctuate over the maximum
at $\chi = 0$. Also, for $|\sigma| \stackrel {_<}{_\sim} M_P$, its
mass is greater than $H$ and, consequently, suffers no quantum
fluctuations in de Sitter space. In hybrid inflation, the choice of
the vacuum is made at the `waterfall' after the end of inflation and,
thus, topological defects can be easily produced.

\par
It is well-known that, as soon as one replaces global by local
supersymmetry, the potential of the theory changes drastically and
inflation becomes, in general, impossible. This is, to a large extent,
due to the generation of a mass for the inflaton which is larger than
$H$. In our model, one can easily check that no such mass is generated
provided we employ the canonical form of the $K\ddot{a}hler$
potential. This fact allows us to hope that, although, our discussion
is not expected to remain unaltered, the modifications necessitated by
the inclusion of  supergravity will not destroy the whole  picture.

\par
To construct an example of a SUSY model with the usual successful MSSM
predictions of $sin^2 \theta_w$ and $\alpha_s$ where our smooth
hybrid inflationary scenario is naturally realized, we consider a model
with gauge symmetry group the subgroup
$G \equiv SU(3)_c \times SU(2)_L \times
SU(2)_R \times U(1)_{B-L} \times U(1)_T $ of $E_6$
and a global symmetry group $C \times B_1 \times B_2
\times R$ commuting with $G$. $C$ is a $Z_2$ matter parity, $B_1$ and
$B_2$ are
$Z_4$ symmetries  and $R$ is a
$U(1) \> R$ -symmetry under which the superpotential $W \rightarrow e^{i
\theta} W$. The coupling constants of the five factors in $G$
are assumed to be equal at a scale $M_c \sim 10^{18} $GeV, probably
related to a more fundamental theory. Let $ q, q^c, \ell, \ell^c, h,
g, g^c,N$ denote chiral superfields transforming under $G$   as
(3,2,1,+1/3,+1), ($\bar{3}$, 1,2,-1/3,+1),  (1,2,1,-1,+1),
(1,1,2,+1,+1),
(1,2,2,0,-2), (3,1,1,-2/3,-2),  ($\bar{3}$,1,1,+2/3,-2) and
(1,1,1,0,+4)
respectively. Also
let $s$ denote
a gauge singlet. The  superfield content of the model consists of three
$q_i$, three $q^c_i$, three $\ell_i$, three $\ell^c_i$, one $h$, one
$\bar {h}$, one $g$, one $\bar{g}$, one $g^c$, one $\bar{g}^c$, one
$\ell^c_o$, one $\bar{\ell}^c_o$, one $\ell_o$, one $\bar{\ell}_o$,
four $q_{(3+m)}$, four $\bar {q}_m$, four $q^c_{(3+m)}$, four
$\bar{q}^c_m$, one $N_o$, one $\bar {N}_o$, three $N_i$, three
$\bar{N}_i$ and one $s$ ($i=1,2,3$ and $m=1,...,4)$. Under $C$, all
superfields remain invariant except $q_i, q_i^c, \ell_i, \ell^c_i$. Under
the generator of $B_1$, all superfields remain invariant except $N_o$ which
changes sign and $\ell^c_o, \bar {\ell}^c_o, g, g^c$ which are
multiplied by $i$. Under the generator of $B_2$, all superfields remain
invariant except $\bar{N}_o$ which changes sign and $\bar{h}, \bar
{g}, \bar {g}^c$ which are multiplied by $i$. Finally under the
$U(1)\> \>
R$ -symmetry, all superfields have charge +1/2 except $N_o, g, g^c$ which
have charge +1/3, $s$ which has charge +1 and $h, \ell^c_o,\bar{\ell}^c_o,
\bar{N}_o$ which have  charge $0$. With the above
charge assignments, the $U(1) \> R$ -symmetry has color anomaly and
therefore is a Peccei-Quinn symmetry solving the strong $CP$ problem.

\par
The gauge group $U(1)_T$ breaks down at a scale $M_X$ through the vevs
of $N_o, \bar{N}_o$. The $D$-flat direction $|N_o| = |\bar{N}_o|$ is
completely $F$-flat as well, because of the global symmetries. The
large Yukawa couplings of the  terms $g g^cN_o, \bar{h}^2 \bar{N}_o$
and $\bar{g} \bar{g}^c \bar{N}_o$ could lead to a radiative change of
the sign of the SUSY breaking mass squared term for one linear
combination of the $N_o$ and
$\bar{N}_o$ bosons at a scale $\sim M_X$ thereby generating a large
vev for this combination  along the $D$-and $F$-flat direction.
The gauge group $SU(2)_R \times U(1)_{B-L}$ breaks down to $U(1)_Y$
also at a scale $M_X$ through the vevs of the neutral components
$\nu^c_o$ and $\bar{\nu}^c_o$ in $\ell^c_o$ and $\bar{\ell}^c_o$. The
relevant superpotential terms are the ones in eq.(1) with the role of
$\phi$ and $\bar{\phi}$ played by $\nu^c_o$ and $\bar{\nu}^c_o$
respectively. The  vev of $N_o$ breaks $B_1$ down  to a $Z_2$
subgroup,
while $\bar{N}_o$ breaks $B_2$ down  to a $Z_2$ subgroup. Besides, the
pair of $<N_o>, <\bar {N}_o>$ breaks the $U(1) \> R$ -symmetry down to a
$Z_2$ discrete symmetry. Starting with $N_o$ and $\bar{N}_o$ at their
vevs we end up with the previously studied cosmological scenario.

\par
We assume that the allowed tree-level mass terms for the gauge
non-singlets are all $\sim M_X$ . Below the scale $M_X$ the only $G_S$
-non-singlet states are the ones of the MSSM. Ordinary light quarks
and leptons are contained in the $q_i,q^c_i, \ell_i, \ell^c_i$ fields
all having negative matter parity. The electroweak Higgs doublets
$h^{(1)}, h^{(2)}$ are the $SU(2)_R$ -doublet partners in $h$ with
mass $\sim  <N^3_o \bar {N}^2_o \nu^c_o \bar {\nu}^c_o > / M^6_c$.
An immediate consequence of this simple
structure is the relation $tan\beta \simeq m_t/m_b$ among the vevs
of $h^{(1)}, h^{(2)}$ and the 3rd generation quark masses. The $g,
g^c$ pair could, in principle, mediate proton decay but the global
symmetries
stabilize the proton almost completely.

\par
Due to the appropriately chosen spectrum, the one-loop renormalization
group equations above $M_X$ predict identical running for the gauge
couplings of all the factors in $G$. Combining this fact with
their assumed equality at $M_c$, we conclude that equality of the
gauge couplings of $G_S$ at  the scale $M_X$ is a justified  boundary
condition. The successful MSSM predictions for $sin^2 \theta_w$ and
$\alpha_s$ follow immediately.

\par
Returning to the cosmologically more interesting neutral sector we
observe that we have exactly three right-handed neutrinos contained in
the three $\ell^c_i$'s which acquire Majorana masses $M_{\nu{^c}}
\sim M^2 \mu^2/M^3_c $ from the non-renormalizable terms $\nu^c_i
\nu^c_j\nu^c_o (\bar{\nu}^c_o)^3 /M^3_c \> \> (i,j=1,2,3)$. Therefore the
$\nu^c_i$ Majorana masses are at most $ \sim 10^{12}$ GeV. The simple
structure of  neutrino Dirac mass terms $m^D_\nu$ leads to the simple
relation $tan \beta \simeq m^D_{\nu_{\tau}} /m_\tau$ for at least the
3rd generation. Thus, we expect a $\nu_\tau$ mass $m_{\nu_{\tau}} \sim
(tan \beta m_\tau)^2/M_{\nu{^c_\tau}}$ which makes the $\nu_\tau$'s a
significant component of the hot dark matter of the universe. The
unbroken $Z_2$ matter parity $C$ guarantees that the lowest
supersymmetric particle remains stable and contributes to the cold
component of the dark matter together with the axion.

\par
The above mentioned terms providing the Majorana neutrino masses
represent the dominant couplings leading to inflaton decay with width
$\Gamma \sim \frac{1}{16 \pi} M^3 \mu^3 M^{-6}_c m$, where $m \sim
10^{14}$ GeV is the inflaton mass. The corresponding `reheat'
temperature $T_r \sim \frac{1}{3} (\Gamma M_P)^{1/2} \sim 10^{11}$
GeV. The $\nu^c_i$'s produced from the inflaton decay generate a
lepton asymmetry$^{(6)}$ which later will be transformed into the observed
baryon asymmetry through the non-perturbative
baryon-and-lepton-number-violating effects in the standard model at
temperatures $\sim 1$ TeV.

\par
The only topological defects predicted by this model are the domain
walls associated with the spontaneous breaking of the discrete
symmetries
$B_1$ and $B_2$ down to their $Z_2$ subgroups.
These symmetries are, at the first place, broken by the vevs of $N_o,
\bar{N}_o$. Our assumption that these fields acquire their vevs before
inflation then leads to the complete absence of catastrophic domain
walls in the universe.

\par
In summary, in the context of SUSY GUTs based on semi-simple gauge groups,
we constructed a smooth version of the hybrid inflationary scenario
which can `naturally' reproduce the observed temperature fluctuations
in CBR  with a GUT scale $\sim 10^{16}$ GeV consistent with the MSSM
unification
and a `compactification' scale
$ \sim 10^{18}$ GeV. This inflationary scenario does not produce any
topological defects. We also presented a specific model where our
smooth hybrid inflation is realized.

\newpage
\section*{References}

\begin{enumerate}
\item A. D. Linde, Particle physics and inflationary cosmology
(Harwood Academic, New York, 1990).

\item A. D. Linde, Phys. Rev. \underline {D49} (1994) 748.

\item G. Dvali, Q. Shafi and R. Schaefer, Phys. Rev. Lett. \underline
{73} (1994) 1886.

\item S. Dimopoulos and H. Georgi, Nucl. Phys. \underline {B193}
(1981) 150;
J. Ellis, S. Kelley and D.V. Nanopoulos, Phys. Lett. \underline {B249}
(1990) 441; U. Amaldi, W. de Boer and H. Furstenau, Phys. Lett.
\underline {B260} (1991) 447; P. Langacker and M. X. Luo, Phys. Rev.
\underline {D44} (1991) 817.

\item A. R. Liddle and D. H. Lyth, Phys. Rep. \underline {231} (1993)
1.

\item G. Lazarides, C. Panagiotakopoulos and Q. Shafi,  Phys. Lett.
\underline {B315} 325; E. \underline {B317} (1993) 661.

\end{enumerate}
\end{document}